\documentclass[prb,reprint,showpacs,amsmath,amssymb]{revtex4-1} 
\usepackage{bm}
\usepackage{graphicx}

\begin{document}

\title{Thermal conductance at the interface between crystals using equilibrium and non-equilibrium molecular dynamics}

\date{\today}

\author{Samy Merabia}\email{samy.merabia@univ-lyon1.fr}
\affiliation{LPMCN, Universit\'e de Lyon; UMR 5586 Universit\'e Lyon 1 et CNRS, F-69622 Villeurbanne, France}

\author{Konstantinos Termentzidis}
\affiliation{1. CETHIL-UMR5008, INSA de Lyon and CNRS, UMR 5008 F-69621 Villeurbanne, France}
\affiliation{2. EM2C UPR CNRS 288 Ecole Centrale Paris, 92295, Ch\^{a}tenay-Malabry, France}

\begin{abstract}
  In this article, we compare the results of non-equilibrium (NEMD) and equilibrium (EMD) molecular dynamics methods to compute the thermal conductance at the interface between solids. 
We propose to probe the thermal conductance using equilibrium simulations measuring the decay of the thermally induced energy fluctuations of each solid. 
We also show that NEMD and EMD give generally speaking inconsistent results for the thermal conductance: Green Kubo simulations probe the Landauer conductance between two solids which assumes phonons on both sides of the interface to be at equilibrium.  On the other hand, we show that NEMD give access to the out-of-equilibrium interfacial conductance consistent with the interfacial flux describing phonon transport in each solid. The difference may be large and reaches typically a factor $5$ for interfaces between usual semi-conductors.
We analyze finite size effects for the two determinations of the interfacial thermal conductance, and show 
that the equilibrium simulations suffer from severe size effects as compared to NEMD. We also compare the predictions of the two above mentioned methods -EMD and NEMD- regarding the interfacial conductance of a series of mass mismatched Lennard-Jones solids. 
We show that the Kapitza conductance obtained with EMD can be well described using the classical diffuse mismatch model (DMM). 
On the other hand, NEMD simulations results are consistent with a out-of-equilibrium generalisation of the acoustic mismatch model (AMM). 
These considerations are important in rationalizing previous results obtained using molecular dynamics, and help in pinpointing the physical scattering mechanisms taking place at atomically perfect interfaces between solids, which is a prerequesite to understand interfacial heat transfer across real interfaces.   
\end{abstract}

\pacs{68.35.Ja, 07.05.Tp, 44.10.+i}

\maketitle
\section{Introduction}
Kapitza conductance controls heat transfer at submicronic length scales in heterogeneous and nanostructured materials. For instance in superlattices, which are made of an arrangement of alternating solid layers, the Kapitza conductance at the interface between the solids controls the overall conductivity of the superlattice when the internal conductance of the solid layers is large
~\cite{chen1998}. Understanding the value of the Kapitza conductance at the interface between solids may thus help in defining directions to minimize or on the contrary maximise the conductivity of the superlattice, with respective applications in energy conversion devices and thermal management. \newline
During the last decade, ultrafast measurements techniques have been developed so that the Kapitza conductance at the interface between a number of metal/dielectrics and dielectrics/dielectrics solids has been characterized~\cite{stoner1993,cahill2003,lyeo2006}. Similarly, ultrafast LASER spectroscopy may also be used to measure the Kapitza conductance between a metal and a solid matrix which can be amorphous~\cite{juve2009}. All the above-mentioned experiments have concluded that the Kapitza conductance is poorly described by the classical AMM and DMM models, with sometimes a difference reaching an order of magnitude. Also the temperature dependence  predicted by the classical models is wrong, with experiments and simulations pointing at a linear increase of the conductance with the temperature~\cite{lyeo2006,hopkins2008,hopkins2009,stevens2007} when the theories predict a constant value at least if interfacial scattering is supposed to be elastic.
These discrepancies may be partly explained by the state of the interface between real materials whose imperfections may
enhance inelastic scattering, thus creating additional energy channels compared with the situation of an ideal interface.  
In this context theoretical modeling may help in pinpointing the physical relevant mechanisms ruling heat transfer across ideal interfaces. To this end, different techniques have been employed including lattice dynamics~\cite{young1989}, Green Function~\cite{zhang2007} and molecular dynamics (MD)~\cite{stevens2007,landry2009}. The latter is a promising method as it is relatively easy to use and it makes no assumption regarding interfacial heat tranport except the classical nature of the energy carriers, a reasonable assumption close to the Debye temperature of the softer solid. However, even for perfect interfaces no agreement has been found between the MD results and the classical AMM and DMM models~\cite{stevens2007,landry2009}. As {\em bulk} transport coefficients, two routes may be followed to determine the interfacial conductance between classical solids~: either the system is driven out-of-equilibrium by creating an interfacial flux using two heat reservoirs on both sides of the interface\cite{schelling2002,stevens2007} or the kinetics of thermally induced fluctuations of the interfacial flux may be recorded around the equilibrium situation where the two solids are at the same temperature~\cite{barrat2003,mcgaughey2006,rajabpour2010}. This latter method relies on the generalisation of the Green-Kubo formulae to interfacial transport coefficients~\cite{barrat2003}. 
Contrary to the case of the thermal conductivity, no agreement has been found between these two methods even when considering simple systems such as the interface between Lennard-Jones solids~\cite{mcgaughey2006}. \newline
In this article, we propose a new method to determine the interfacial conductance in the spirit of the EMD method using the energy autocorrelation function of each solid on both sides of the interface. This may solve practical problems frequently encountered in equilibrium simulations when a plateau in the integral of the relevant correlation function should be identified, which often leads to practical difficulties.
We explain the discrepancies between the EMD and the NEMD simulations determination of the interfacial heat conductance. We show that the EMD yields the {\em Landauer} conductance which assumes phonons on both sides of the interface to have equilibrium distribution. On the other hand, we will show that the conductance measured in NEMD is well described by the general expression of Simons which accounts for the out-of-equilibrium phonon distribution consistent with the created heat flux\cite{simons1974,katerberg1977}. Thus we conclude that the two methods give {\em intrinsically} inconsistent values of the interfacial conductance. The difference is important and may reach nearly an order of magnitude for solids displaying moderate acoustic mismatch. We analyze also the finite size effects in the two methods and show that EMD suffers from stronger size effects than NEMD. Finally, we analyze the interfacial conductance at the interface between a series of mass-mismatched Lennard-Jones solids using both methods. We show that the classical DMM model provides a good description of the EMD data. On the other hand, both the AMM and DMM models fail to predict the conductance obtained in NEMD. A good agreement is found if we extend the AMM model by accounting for the out-of-equilibrium phonon distribution consistent with the imposed interfacial flux. \newline 
The article is structured as follows: in the section~\ref{theory}, we first review the basics of interfacial heat transport; We discuss the difference between the Landauer conductance which assumes the energy carriers to be described locally by equilibrium distribution functions and the general expression proposed by Simons. In the section~\ref{Green-Kubo}, we show
the connection between the Landauer conductance and the decay of the energy autocorrelation function in each solid. This allows us to propose an alternate expression to measure the interfacial conductance using EMD. This methodology is applied in the section~\ref{simulations} where we analyze the case of the interface between Lennard-Jones solids having a variable mass contrast. We also compare the conductance obtained using the two methods with the different theoretical predictions discussed in the section~\ref{theory}. We discuss the consequences of this work in the Conclusion.          

\section{Theory \label{theory}}
In this section, we briefly review the basic definitions of the interfacial conductance and we discuss its relation with the phonon distribution on both sides of the interface. The equations derived in this section are not completely new but they are reviewed for the sake of completeness. In particular, we review the expression first proposed by Simons~\cite{simons1974} which accounts for the out-of-equilibrium phonon distribution consistent with the interfacial flux.\newline 

\begin{figure}
\includegraphics[width=0.9\linewidth]{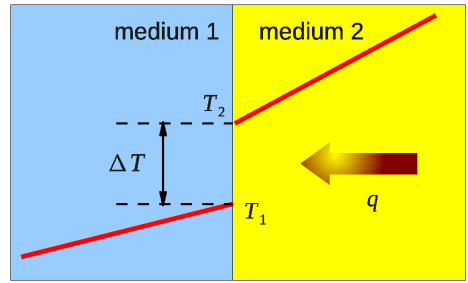}
\caption{(Color online) Temperature jump across the interface between two media crossed by an interfacial flux $q$. The temperature profile in each medium is schematically represented by red solid lines.}
\label{fig_temperature_jump}
\end{figure}

\begin{figure}
\includegraphics[width=0.9\linewidth]{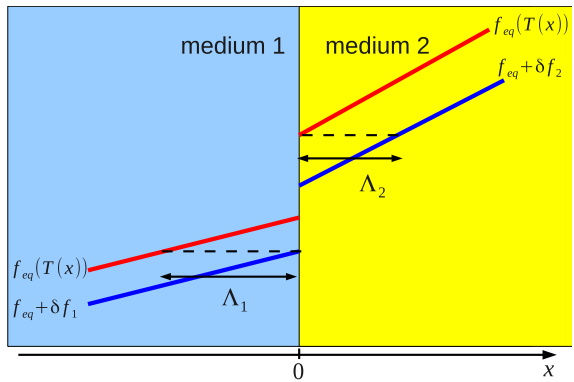}
\caption{(Color online) Steady phonon distribution function on both sides of the interface crossed by a heat flux.
Red solid lines represent the local equilibrium distribution function $f_{\rm eq}(T(x))$ given by the Bose-Einstein distribution eq.~(\ref{Bose_Einstein}). We have considered classical phonons for which the Bose-Einstein distribution is proportional to the temperature. Blue solid lines represent the out-of-equilibrium distribution $f_{\rm eq}(T(x))+\delta f$ 
of incident phonons where $\delta f$ is given by eq.~(\ref{deviation_distribution}). We have considered a phonon mode propagating in the direction of the temperature gradient so that $\delta f<0$.
We have also illustrated the graphical construction inherent to the definition of the equivalent equilibrium temperature (see the eqs.~\ref{def_equivalent_temperature}
and \ref{equivalent_temperature}). $\Lambda_1$ and $\Lambda_2$ are the phonon mean free paths of the considered phonon mode in each medium.}
\label{fig_equivalent_temperature}
\end{figure}
Consider the interface between two media $1$ and $2$ as sketched in fig.~\ref{fig_temperature_jump}. The interfacial conductance $G$ between these two media is defined in terms of the ratio
\begin{equation}
\label{definition_conductance}
G= q/(T_2-T_1)
\end{equation} 
where $q$ is the heat flux flowing across the interface from medium $2$ to $1$, and $T_i$ denotes the temperature of the medium $i$ 
in the vicinity of the interface. The interfacial conductance eq. \ref{definition_conductance} can be related to the phonon distribution in each medium if the heat flux $q$ is expresed in terms of transmitted phonons~:
\begin{eqnarray}
\label{interfacial_flux}
q=\frac{1}{V}\sum_{p,\vec k}^{+} v_{1x}(p,\vec k) \hbar \omega (p, \vec k) f_1(p,\vec k) t_{12}(p,\vec k) \nonumber \\
+\frac{1}{V} \sum_{p,\vec k}^{-} v_{2x}(p,\vec k) \hbar \omega (p,\vec k)f_2 (p,\vec k) t_{21}(p,\vec k)
\end{eqnarray} 
where $V$ is the volume of each medium supposed to be equal, $v_{ix}$ is the group velocity in medium $i$ projected along the direction $x$ normal to the interface, $f_i$ is the mode dependant phonon distribution function in medium $i$, $t_{ij}(\vec k)$ is the wave vector dependant transmission coefficient from medium $i$ to medium $j$, and the sums runs over all polarizations indexed by $p$, and over wavevectors in the first Brillouin zone corresponding to phonons crossing the interface, {\em i.e} those for which $v_{1x}>0$ and $v_{2x}<0$ respectively. In the following we will drop the variables $(p,\vec k)$ indexing the phonon polarization and wavevector
to simplify the notations. We will refer to any quantity depending on $(p,\vec k)$ as mode-dependent. \newline 
The problem of the determination of the interfacial conductance eq.\ref{definition_conductance} relies on our knowledge of the 
phonon distribution functions $f_i$ at both sides of the interface. The simplest reasoning is to assume that the phonons population can be described by the equilibrium distribution $f_{eq}$ given by the Bose-Einstein distribution:
\begin{equation}
\label{Bose_Einstein}
f_{\rm eq}(\omega,T)= \frac{1}{\exp \left( \hbar \omega / k_B T \right) -1 }
\end{equation}
at the temperature $T_i$ in the vicinity of the interface. The two sums appearing in eq.~\ref{interfacial_flux} can be contracted to a single sum on the phonon population coming from medium $1$ if we invoke the principle of detailed balance 
in the situation where the two media are at thermal equilibirum at the common temperature $T_1$
 when the flux $q$ vanishes~\cite{landry2009}. One arrives then at the Landauer formula for the interfacial conductance~\cite{landauer1970}:
\begin{equation}
\label{Landauer}
G_{\rm eq} = \frac{1}{V} \sum_{p,\vec k}^{+} \hbar  \omega v_{1x} t_{12} \frac {\partial f_{eq}} {\partial T}  
\end{equation}
Note of course that using the principle of detailed balance, the Landauer conductance can be expressed as a function 
of the transport properties characterizing the medium $2$:
\begin{equation}
G_{\rm eq} = \frac{1}{V} \sum_{p,\vec k}^{-} \hbar  \omega v_{2x} t_{21} \frac {\partial f_{eq}} {\partial T}  
\end{equation}
The Landauer formula has commonly been used in the determination of the Kapitza conductance \cite{young1989,pettersson1990,swartz1989}.
Its limitations are well known \cite{simons1974,katerberg1977,pettersson1990,chen2003,landry2009}~: equation~(\ref{Landauer}) predicts a finite conductance when the two materials are identical, {\em i.e.} when $\forall \vec k, t_{12}(\vec k)=1$, which is of course contrary to the intuition:  for an interface between similar media, the temperature drop should vanish whatever the flux $q$, leading to an infinite conductance~\cite{note_infinite_conductance}.  Obviously, the problem is related to the use of two equilibrium distribution functions in the flux eq.\ref{interfacial_flux}. The previously mentioned paradox may be solved using the actual distribution function consistent with the interfacial heat flow. This analysis has been done by Simons~\cite{simons1974} and generalized by
Chen\cite{chen2003} and Landry and McGaughey\cite{landry2009}. The out of equilibrium distribution function is supposed to obey the Boltzamnn transport equation (BTE) under the relaxation time approximation~\cite{ziman2001}:
\begin{equation}
\label{BTE}
\frac{\partial f_i}{\partial t} + \vec v_i \cdot \vec \nabla f_i = -\frac {f_i-f_{\rm eq}}{\tau_i(\omega)}  
\end{equation}
where $f_{\rm eq}$ is the Bose-Einstein distribution given in eq.~\ref{Bose_Einstein} and $\tau_i$ is the mode dependant relaxation time supposed to depend only on the frequency $\omega$. In steady state, a solution of the BTE equation eq.~\ref{BTE} can be found in the form: 
\begin{equation}
\label{steady_state_distribution}
f_i(\vec r)=f_{\rm eq}(T(\vec r))+\delta f_i(\vec r)
\end{equation}
where $T(\vec r)$ is the local value of the temperature.
We assume in this way that the temperature is defined at any point of the material, an assumption which is reasonable if the phonon mean free path in each medium is not too large compared to the characteristic dimensions of the system, {\em i.e.} the distance between the interface and the heat reservoirs. Anyway, from a practical point of view in MD, one can always think of the local temperature as the mean kinetic energy of the atoms in a small volume encompassing the point $\vec r$. If furthermore, we assume that in each medium the temperature profile is linear, an assumption which is again confirmed by NEMD simulations~\cite{note2}, then the deviation from the local equilibrium writes:
\begin{equation}
\label{deviation_distribution}
\delta f_i(\vec r) = -\tau_i \frac{\partial f_{\rm eq}}{\partial T} \vec v_i \cdot \vec \nabla T
\end{equation}
Hence, the excess of phonons propagating in each medium is proportional to the heat flux. Phonons travelling in the direction of the flux are in excess while phonons travelling in the opposite direction are depleted. A schematic representation of the distribution of incident phonons across the interface is displayed on fig.~\ref{fig_equivalent_temperature}.
Injecting the latter distribution function given by eqs.~\ref{steady_state_distribution} and \ref{deviation_distribution}, 
in the interfacial flux $q$ eq.~\ref{interfacial_flux}, one arrives at:
\begin{eqnarray}
q=G_{\rm eq} (T_2-T_1) - \sum_{p,\vec k}^{+} \tau_1 v_{1x}^2 \hbar \omega \frac{\partial f_{\rm eq}}{\partial T} t_{12} \frac {\partial T}{\partial x} \vert_1  \nonumber \\
- \sum_{p,\vec k}^{-} \tau_2 v_{2x}^2 \hbar \omega \frac{\partial f_{\rm eq}}{\partial T} t_{21} \frac {\partial T}{\partial x} \vert_2
\end{eqnarray}
where the temperature gradients are estimated on both sides of the interface. The two temperature gradients can be eliminated if we assume diffusive heat transport in each 
medium so that $q=-\lambda_1 \frac{\partial T}{\partial x} \vert_1=-\lambda_2 \frac{\partial T}{\partial x} \vert_2$ where $\lambda_i$ denotes the thermal conductivity of medium $i$. The interfacial conductance writes then:
\begin{equation}
\label{neq_conductance}
G_{\rm neq}=\frac{G_{\rm eq}}{1-\beta_{12}-\beta_{21}}
\end{equation}
where we have introduced the fractions:
\begin{equation}
\label{beta}
\beta_{12} = \frac{1}{V} \sum_{p,\vec k}^{+}   \tau_1 v_{1x}^2 \hbar \omega \frac{\partial f_{\rm eq}}{\partial T} t_{12}/\lambda_1
\end{equation}
and a similar equation for $\beta_{21}$. The physical signification of $\beta_{12}$ is clear~: it is a measure of the fraction of the energy flux flowing across the interface that is transmitted. This coefficient varies typically between $0$ when all the phonon modes of medium $1$ are reflected by the interface to $1/2$ when all the modes of medium $1$ are transmitted. 
In particular, if we consider the case of similar materials, it is easy to show that 
the interfacial conductance eq.~\ref{neq_conductance} diverges to infinity using the Peierls expression for the 
thermal conductivity~\cite{ziman2001}:
\begin{equation}
\label{conductivity_Peierls}
\lambda_1 = \frac{1}{V} \sum_{p,\vec k} \tau_1 v_{1x}^2 \hbar \omega \frac{\partial f_{\rm eq}}{\partial T} \rightarrow \beta_{12}(t_{12}=1)=\frac{1}{2}
\end{equation}
where the last equality applies to the case of an interface which transmits all the phonon modes, $\forall (p,\vec k), t_{12}=1$.
Thus at least, eq.~\ref{neq_conductance} solves the paradox of the conductance of the interface between identical materials. 
Note that we could have obtained the same expression for the conductance using the concept of equivalent {\it equilibrium} temperatures.
By definition, the equivalent equilibrium temperature may be defined in a classical system in terms of the kinetic energy of the incident phonons in the vicinity of the interface. This condition is graphically illustrated in fig.~\ref{fig_equivalent_temperature} and is mathematically expressed by~:
\begin{equation}
\label{def_equivalent_temperature}
f_{\rm eq}(T_1^{\rm eq}) = f_{\rm eq}(T_1) + \delta f_1
\end{equation}
yielding 
\begin{equation}
\label{equivalent_temperature}
T_1^{\rm eq}=T_1 - \Lambda_1 \cos \theta \frac{\partial T}{\partial x}
\simeq T(x=-\Lambda_1 \cos \theta)
\end{equation}
 where the interface is supposed to be localized at $x=0$. Here the $\theta$ is the angle of incidence and $\Lambda_1$ is the mean free path of the considered phonon mode. Thus, we have shown that the equivalent equilibrium temperature is the temperature of incident phonons at a distance of one mean free path away from the interface as exemplified in fig.~\ref{fig_equivalent_temperature}. This explains why Aubry et {\em al.} obtained an expression similar to Simon conductance using the equilibrium distribution of phonons at a distance one mean free path from the interface~\cite{pettersson1990,aubry2008}. Note that in the previous discussion and in the formula used by Aubry et al., the equivalent temperature is a mode dependent quantity, as both $\theta$ and $\Lambda_1$ depend on the considered mode.
Using the concept of equivalent temperatures may be dangerous because one may be tempted to believe that the phonon population is at equilibrium at a distance $\Lambda$ away from the interface, which is of course wrong. It is nevertheless not surprising to find the same value of the interfacial conductance using the concept of equivalent temperature at a distance $\Lambda$, because the effective incident flux that may be transmitted by the interface comes from phonons which have not been scattered by other phonons before reaching the interface~\cite{pettersson1990} and as a first approximation if temperature gradients are not too large the corresponding phonon population may be described by $f_{\rm eq}(x=- \Lambda \cos \theta)$. \newline 
In the following, it will be useful to express the different conductances in terms of the vibrational density of states (vDOS):
\begin{equation}
\label{VDOS}
g_{p}(\omega)=\frac{1}{V} \sum_{\vec k} \delta (\omega -\omega_{p,\vec k}) 
\end{equation}
where the sum runs over the eigenmodes of the crystal in the first Brillouin zone. In the common case where the transmission coefficients depend only on the frequency $\omega$ and on the incident angle $\theta$, the Landauer conductance is:
\begin{eqnarray}
\label{Landauer_g}
G_{\rm eq}=\frac{1}{2}\sum_p \int_{0}^{\omega_{\rm max}} g_{1,p}(\omega)\vert v_1(\omega) \vert \hbar \omega \frac{\partial f_{\rm eq}}{\partial T} \times \nonumber \\ \int_{0}^{\pi/2}t_{12}(\omega,\theta) \cos \theta \sin \theta d \theta d\omega
\end{eqnarray}  
and the fraction $\beta_{12}$ becomes:
\begin{widetext}
\begin{equation}
\label{beta/lambda}
\beta_{12}=\frac{\frac{1}{2}\sum_p \int_{0}^{\omega_{\rm max}} g_{1,p}(\omega) \tau_1(\omega) \vert v_1(\omega) \vert^2 \hbar \omega \frac{\partial f_{\rm eq}}{\partial T} \int_{0}^{\pi/2}t_{12}(\omega,\theta) \cos^2 \theta \sin \theta d \theta d\omega}{\frac{1}{3}\sum_p \int_{0}^{\omega_{D,1}} g_{1,p}(\omega) \tau_1(\omega) \vert v_1(\omega) \vert^2 \hbar \omega \frac{\partial f_{\rm eq}}{\partial T} d\omega}
\end{equation}                                                                                                                                                                                                                                                                                                                                                                                                                                                                                                                                                                                     
\end{widetext}
where the factor $1/2$ in the numerator comes from the integration over the azimuthal angle $\phi$ and the integration is carried out 
over the first Brillouin zone. $\omega_{\rm max}$ is the maximal frequency transmitted by the interface and its value will be discussed later and $\omega_{D,1}$ is the Debye frequency in medium $1$.
Again, a similar expression for the term $\beta_{21}$ can be obtained by permuting in the previous equation the indexes $1$ and $2$.
The challenge is now to specify the lifetimes $\tau_i(\omega)$ and the transmission coefficients. 
We will discuss possible expressions for $t_{12}$ based on traditional interfacial transport models in the section \ref{conductance_series_massmismatch} when we will analyze the conductance obtained by NEMD. \newline
So far, we have seen two formulae relating the interfacial thermal conductance to the energy transmission coefficient:
the Landauer fomula~eq.~\ref{Landauer} which assumes that the phonons on both sides of the interface are at equilibrium,
and the general formula eq.~\ref{neq_conductance} which accounts for the actual out of equilibrium distribution of the 
phonons in the vicinity of the interface. Now the question that we want to answer is: what do we measure in a molecular dynamics simulation? Intuitively, in NEMD simulations where the system is crossed by a flux, we should measure a conductance given by eq.~\ref{neq_conductance} because the system is subject to a large temperature gradient (on the order of $1$ K/nm !) 
and the phonons can not be considered locally at equilibrium. On the other hand, it seems reasonable to consider that in an equilibrium 
simulations where thermally induced fluctuations of the interfacial flux are probed, one should measure the Landauer conductance 
eq.~\ref{Landauer} rather than the non-equilibrium conductance eq.~\ref{neq_conductance}.
We will make this point more quantitative in the next section.

\section{Green-Kubo formulae: conductance from equilibrium fluctuations \label{Green-Kubo}}
In this section, we derive Green-Kubo formulae for the interfacial conductance.
We will prove that the Puech formula traditionally used in equilibrium simulations to measure the interfacial conductance 
is exactly given by the Landauer conductance eq.~(\ref{Landauer}) and thus differs from the non-equilibrium conductance 
$G_{\rm neq}$ (eq.(\ref{neq_conductance})). We will also propose an equivalent formula easier to evaluate in molecular simulations. \newline
The general idea behind Green-Kubo formulae is that the regression of the fluctuations of an internal variable obeys macroscopic laws. 
In the case of interfacial heat transfer, the relevant variable is the interfacial flux $q$ and the corresponding Green-Kubo formula reads:
\begin{equation}
\label{Puech_formula}
G_{\rm Puech} = \frac{1}{\mathcal A k_B T^2} \int_{0}^{+\infty} \langle q(t) q(0) \rangle dt
\end{equation}
This formula has been used for solid/liquid interfaces~\cite{barrat2003} and superlattices as well~\cite{mcgaughey2006,chalopin2012}. 
However practically in a MD simulation, the expression of the heat flux $q$ involves only atoms near the interface~\cite{barrat2003} and it is also sometimes difficult to estimate the 
plateau in the heat flux correlation function in eq.~\ref{Puech_formula}. In the following, we will show that we can improve the statistics on the determination of the interfacial conductance by measuring the fluctuation of the mechanical energy of each solid 
on both sides of the interface. In passing, we will show that for the case of solid/solid interfaces, the Puech formula
eq.~(\ref{Puech_formula}) identifies with the Landauer conductance eq.~(\ref{Landauer}).\newline  
To this end, we consider two semi-infinite media separated by an interface whose area is denoted $\mathcal A$. The two media are supposed to be at thermal equilibrium at the same temperature $T$, and the energy in each medium can change only because of exchange of energy with the other medium through the interface. Generally speaking, the energy fluctuation in each medium is~\cite{stephenson1983}: $\langle \delta E_i^2 \rangle = k_B T^2 /(1/C_{v1}+1/C_{v2})$ where $E_i(t)$ is the instantaneous mechanical energy of the medium $i$, and $C_{v1}, C_{v2}$
are the specific heat characterizing the two media. 
Note that the relevant statistical ensemble to describe the fluctuations of $E_i$ is neither $NVE$ because only the total energy $E_1+E_2$ is conserved nor $NVT$ because stricly speaking each system is not in contact with a thermostat but with a system which is comparable in size. We refer the reader to Stephenson~\cite{stephenson1983} for a derivation of the fluctuations of the different 
quantities in this situation.
The classical $NVT$ formula is however recovered when one of the two media (say $2$) has a large number of degrees of freedom
so that $C_{v2} \gg C_{v1}$. In the following, we will assume that the two media have the same specific heat so that the energy flucuation in each medium is: 
\begin{equation}
\label{energy_fluctuation}
\langle \delta E_i^2 \rangle = k_B T^2 C_v/2
\end{equation}
 This hypothesis will not affect the 
final result but allows to simplify the notations all along the derivation. 
The fluctuations of the interfacial flux $q$ are related to the fluctuations of the energy in medium $1$ through the energy conservation 
equation:
\begin{equation}
\label{energy_conservation}
\frac{d E_1}{dt}=-q \mathcal A
\end{equation}
where $q$ is the instantaneous value of the interfacial energy flux flowing from the medium 1 towards medium 2,
which in the situation considered fluctuates around zero. This flux may be expressed in terms of excess phonon occupation number
$\delta n_i$:
\begin{equation}
q=\frac{1}{V} \sum_{\vec k}^{+} v_{1x} \hbar \omega \delta n_1 t_{12} + \sum_{\vec k}^{-} v_{2x} \hbar \omega \delta n_2 t_{21}
\end{equation}
where the excess phonon occupation number is simply related to the fluctuation of the energy: $\delta E_i = \sum_{\vec k} \hbar \omega 
\delta n_{i,\vec k}$ and here $\vec k$ is a shorthand notation representing the wavevector and the polarization. In the following, it will be useful to rewrite the energy conservation:
\begin{equation}
\label{derivative_energy}
\frac{d E_1}{dt}=-\frac{1}{V} \sum_{\vec k}^{+} \frac{\hbar \omega \delta n_1}{\tau_{\vec k}}
 + \frac{1}{V} \sum_{\vec k}^{-} \frac{\hbar \omega \delta n_2}{\tau_{\vec k}}
\end{equation}
where we have introduced the mode dependent relaxation times:
\begin{eqnarray}
\label{relaxation_time}
\tau_{\vec k}&=& {V}/{\mathcal A v_{1x} t_{12}} \; \mathrm{if} \; v_{1x}>0 \quad \text{(first sum in the rhs of eq.~\ref{derivative_energy})} \nonumber \\
             &=& {V}/{\mathcal A \vert v_{2x} \vert t_{21}} \; \mathrm{if} \; v_{2x}<0 \quad \text{(second sum)}
\end{eqnarray}
These relaxation times may be interpreted as interfacial scattering terms and are independent of the bulk phonon relaxation times.
If we assume the different modes to be independent and consistently with eq.~(\ref{energy_fluctuation}) characterized by a variance 
\begin{equation}
\label{fluctuation_occupation_number}
\langle \delta n_{i,\vec k}^2 \rangle = \frac{k_BT^2 \bar c_v}{2 \hbar \omega_{\vec k}} 
\end{equation}
where $\bar c_v=\hbar \omega_{\vec k} \frac{\partial f_{\rm eq}}{\partial T}$ is the mode dependent specific heat,
the energy autocorrelation function follows~:
\begin{equation}
\label{energy_correlation}
\langle \delta E_1(t) \delta E_1(0) \rangle = \sum_{\vec k} \frac{k_B T^2 \bar c_v}{2} \exp \left(-\vert t \vert/\tau_{\vec k}\right) 
\end{equation}
where we have used the total energy conservation $\delta E_1 = -\delta E_2$.
Differentiating this latter equation, one arrives at: 
\begin{eqnarray}
\langle \frac{d E_1(t)}{dt}  \delta E_1(0) \rangle & = & -\sum_{\vec k} \frac{k_B T^2 \bar c_v}{2 \tau_{\vec k}} \text{sgn}(t) \exp \left(-\vert t\vert/\tau_{\vec k} \right) \nonumber \\ 
& + &  \sum_{\vec k} k_B T^2 \bar c_v \delta(t)
\end{eqnarray}
where $\text{sgn}(t)$ is the sign function and the second term in the right hand side comes from the discontinuity of the derivative 
of $\exp(-\vert t \vert/\tau_{\vec k})$ at the origin~\cite{note_derivative}.
The sums over all the wavevectors $\vec k$ may be transformed in a sum running over the modes crossing the interface if we express 
the detailed balance condition:
\begin{equation}
\label{detailed_balance}
\bar c_v v_{1x} t_{12} \vert_{v_1x>0}=  -\bar c_v v_{2x} t_{21} \vert_{v_2x<0}
\end{equation}
yielding for $t>0$:
\begin{eqnarray}
\langle \frac{d E_1(t)}{dt}  \delta E_1(0) \rangle & = & -\frac{1}{V} \sum_{\vec k}^{+} \mathcal A k_B T^2 \bar c_v t_{12} v_{1x} \exp \left(-t/\tau_{\vec k} \right) \nonumber \\ 
& + & 2 \sum_{\vec k}^{+} k_B T^2 \bar c_v \delta(t)
\end{eqnarray}
Note that in the thermodynamic limit the first term vanishes. For a finite system and when the time $t \rightarrow 0^{+}$, 
the second term involving a Dirac distribution may be neglected and one has~:
\begin{equation}
\label{energy_Landauer}
-\frac{1}{\mathcal A k_BT^2} \left( \frac{d C_{EE}}{dt} \right)_{t=0^{+}} = \frac{1}{V} \sum_{\vec k}^{+} \bar c_v t_{12} v_{1x} = G_{\rm eq}
\end{equation}
where $C_{EE}(t)=\langle \delta E_1(t) \delta E_1(0) \rangle $ is the energy auto-correlation characterizing solid $1$ and $G_{\rm eq}$ is the Landauer conductance defined in eq.~\ref{Landauer}. 
The previous equation eq.~(\ref{energy_Landauer})
is a new Green-Kubo formula for the interfacial conductance which relates the slope of the energy autocorrelation function at the origin to the Landauer conductance.  We will show in the next section that this formula may be easier to estimate in a MD simulation than the classical Puech formula which requires to 
identify a plateau in the running integral of the heat flux autocorrelation function. Alternately, we can also relate the Puech formula to the Landauer conductance in the thermodynamic limit by remarking that 
$\langle \frac{d \delta n_{1,\vec k}(t)}{dt} \frac{d \delta n_{1,\vec k}(0)}{dt} \rangle = -\frac{d^2}{dt^2} \langle \delta n_{1,\vec k}(t) \delta n_{1, \vec k}(0) \rangle$ to arrive at
\begin{eqnarray}
\langle \frac{d E_1(t)}{dt} \frac{d E_1(0)}{dt} \rangle & = & -\sum_{\vec k}^{+} \frac{k_B T^2 \bar c_v}{\tau_{\vec k}^2} \exp
\left(-\vert t \vert/\tau_{\vec k} \right) \nonumber \\
& + & 2\frac{\mathcal A k_BT^2}{V} \sum_{\vec k}^{+} \bar c_v t_{12} v_{1x} \delta(t)
\end{eqnarray}
In the thermodynamic limit, in principle the first term $\propto \mathcal A  (\mathcal A/V)$~\cite{note_thermodynamic_limit} is negligible compared with the second $\propto \mathcal A$, and one has the following Green-Kubo equation:
\begin{equation}
\label{Green_Kubo_infinite_system}
\frac{1}{\mathcal A k_BT^2} \int_{0}^{+\infty} \langle \frac{d E(t)}{dt} \frac{d E(0)}{dt} \rangle dt = G_{\rm eq}
\end{equation}
This equation is exactly the Puech formula used to calculate the liquid/solid Kapitza resistance~\cite{puech1986,barrat2003}. 
We have shown that for solids/solids this formula identifies with the Landauer conductance. It is important to realize that 
the previous formula has been derived for an infinite system size.
For a finite system on the other hand, the running integral 
\begin{eqnarray}
\label{Green_Kubo_finite_system}
\frac{1}{\mathcal A k_BT^2} \int_{0}^{t} \langle \frac{d E(t')}{dt} \frac{d E(0)}{dt} \rangle dt' = \frac{1}{V} \sum_{\vec k}^{+} 
\bar c_v t_{12} v_{1x} \nonumber \\
 - \frac{1}{V} \sum_{\vec k}^{+} \bar c_v t_{12} v_{1x} \left(1 -\exp \left(-t/\tau_{\vec k}\right)\right)
\end{eqnarray}
will consist of two parts: the first term is the Landauer conductance, the second term is negative and corresponds to the final decay of the running integral. \newline 
In the next section, we will compare the formulae eqs.~(\ref{energy_Landauer}) and (\ref{Puech_formula}) to the results of NEMD simulations.

\section{Simulations \label{simulations}}
\subsection{Lennard-Jones systems}
We now study how the previous formulae may be used in molecular dynamics simulations to estimate the Kapitza conductance between two solids. All the following results have been obtained for the case of the interface between Lennard-Jones solids. There are numerous advantages 
to work with LJ solids. The first is the simplicity of the interaction potential as compared to many body potentials used to model
semi-conductors. This has an important practical consequence as it allows to run simulations with large system lengths because 
of the relatively short computational times required. Also, from a thermal point of view there is no need to worry about optical phonon modes.

\subsection{Structures}
\begin{figure}
\includegraphics[width=0.9\linewidth]{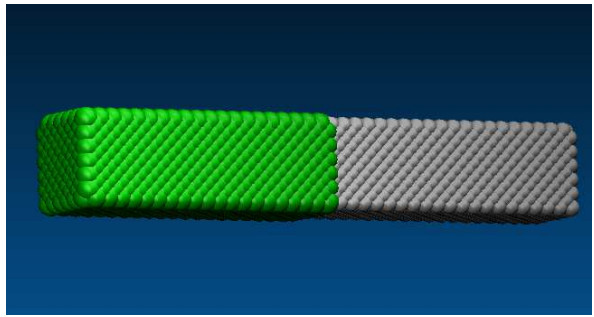}
\caption{(Color online) Configuration studied: interface separating two Lennard Jones fcc solids having the same lattice constant but different masses.}
\label{structure_LJ_interface}
\end{figure}

We will consider systems consisting of two perfect fcc Lennard-Jones solids whose interface is orientated along the crystallographic [100] direction. The section of the system is fixed to $6 a_0 \times 6 a_0$ where $a_0$ is the fcc lattice constant, and the thickness of each medium has been varied between $10$ and $50$ $a_0$. A typical initial configuration is represented in figure~\ref{structure_LJ_interface}. 
All the atoms of the system interact through a Lennard-Jones potential $V_{\rm LJ}(r) = 4\epsilon \left((\sigma/r)^{12}-(\sigma/r)^6\right)$ truncated at a distance $2.5 \sigma$. A single set of energy $\epsilon$ and diameter $\sigma$ characterizes the interatomic interaction potential. As a result, the two solids have the same lattice constant $a_0$, and the interface may be considered perfect.
To introduce an acoustic mismatch between the two solids, we have considered a mass mismatch between the masses of the atoms of the two solids, characterized by the mass ratio $m_r =m_2/m_1$, which will take typical values between $1$ and $10$. From now on, we will use real units where $\epsilon=1.67 \; 10^{-21}$ J; $\sigma =3.4 \; 10^{-10}$ m and $m_1=6.63 \;10^{-26}$ kg, where these different values have been chosen to represent solid Argon. With this choice of units, the unit of time is $\tau=\sqrt{m\sigma^2/\epsilon}=2.14$ ps, the unit of thermal conductivity is $\lambda= \frac{k_B}{\sigma^2}\sqrt{\epsilon/m} \simeq 18.8$ $10^{-3}$ $W/K/m$ and the unit of interfacial conductance is $G = k_B/(\tau \sigma^2) \simeq 56 MW/K/m^2$. The different interfaces have been prepared as follows: 
first the structures have been generated by mapping the space with fcc structures using the lattice parameter of the fcc LJ solid at zero temperature~\cite{chantrenne2003}: $a_0(T=0 K)= 1.5496 \sigma$. The structures have been then equilibrated at the final finite temperature $T=40$ K using first a Berendsen thermostat and a barostat at $0$ atm~\cite{frenkelsmit}. 
Once the instantaneous temperature has increased to a value close to the final expected temperature, we have switched off the Berendsen thermostat and used a Nos\'e Hoover thermostat. The total equilibration time lasts one million time steps which correspond to a total time of $4,28$  ns. All the systems studied have been equilibrated at the temperature $T=40$ K, and the lattice parameter at this temperature has been found to be: $a_0=1.579 \sigma$. In EMD, periodic boundary conditions have been applied in all spatial directions so that the system represented is a superlattice~\cite{chalopin2012}. On the other hand in NEMD we use periodic boundary conditions only in the directions parallel to the interface. 
%

\subsection{Computing the interfacial conductance with NEMD}
\begin{figure}
\includegraphics[width=0.9\linewidth]{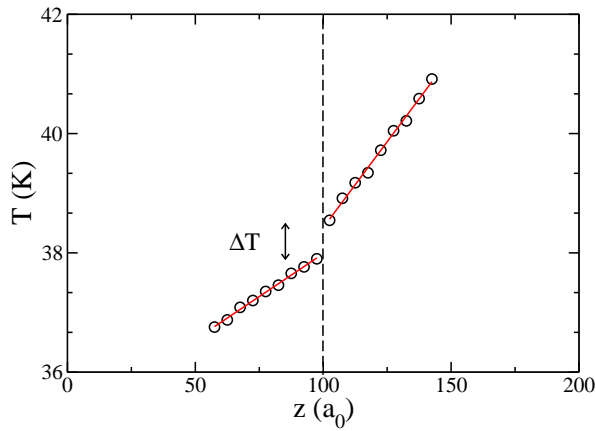}
\caption{(Color online) Stationary temperature profile across the interface obtained by NEMD. The symbols are the simulation data while the solid lines are the linear extrapolation used to compute the interfacial temperature drop $\Delta T$. The position of the interface is located by the vertical dashed lines.\label{temperature_profile}}
\end{figure}

Alternately, we will compare the results of EMD to NEMD. The principle of these latter simulations has been already described elsewhere \cite{landry2009,termentzidis2011} and we just focus here on the details 
of the technique. We impose a thermal flux perpendicular to the interface between the two solids by thermostatting in each medium 
 two layers of atoms remote from the central interface at the respective temperature $T_C=40-3.6$K and $T_H=40+3.6$ K, while the end atoms are maintained fixed.
The size of the cold and hot regions has been found to have negligible effect on the measured conductance.
After a number of time steps varying between $500000$ for the smallest system to $5$ million for the largest, we monitor the 
temperature profile in each medium using the kinetic energy of the particles. 
A typical example of the corresponding temperature profile is shown in figure~\ref{temperature_profile} zooming on the vicinity of the interface.  The interfacial conductance is obtained from the heat flux and the temperature jump across the interface, these latter quantities being measured using the heat power delivered by the heat source which is monitored during several million of time steps. The temperature jump $\Delta T$ is obtained by extrapolation of the linear profiles in the two media as shown in figure \ref{temperature_profile}. In the following, we will present results for the interfacial conductance obtained using $5$ independent simulations.
%

\subsection{Computing the interfacial conductance with EMD}
\begin{figure}
\includegraphics[width=0.9\linewidth]{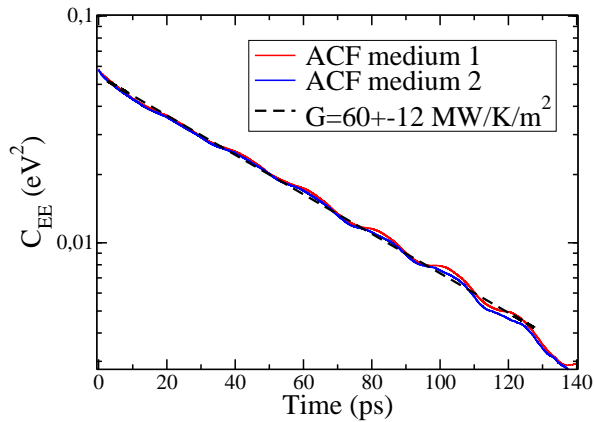}
\caption{(Color online) Energy autocorrelation functions of the two $fcc$ Lennard-Jones crystals separated by a planar interface 
obtained with molecular dynamics simulations. Dashed lines show the exponential fit. 
The parameters are: Total Length=$40$ $a_0$; $T=40$ K; mass ratio $mr=2$.}
\label{EACF}
\end{figure}

\begin{figure}
\includegraphics[width=0.9\linewidth]{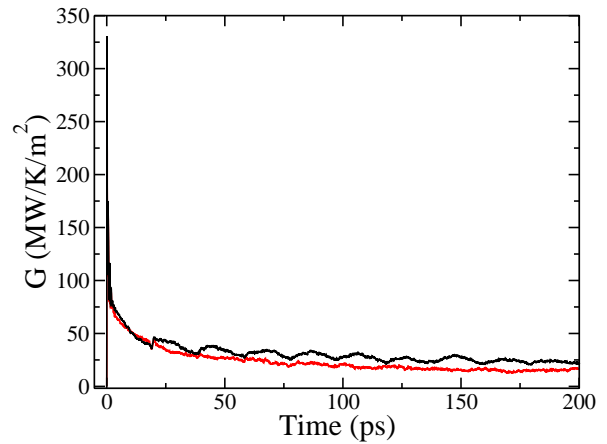}
\caption{(Color online) Interfacial conductance of the same interface as fig. \ref{EACF} calculated using the 
Puech Green-Kubo formula eq.~\ref{Puech_formula}. The two curves correspond to the two interfaces of the system.
Each curve is an average over $30$ independant trajectories. Same parameters as figure \ref{EACF}.}
\label{fig4}
\end{figure}

\begin{figure}
\includegraphics[width=0.9\linewidth]{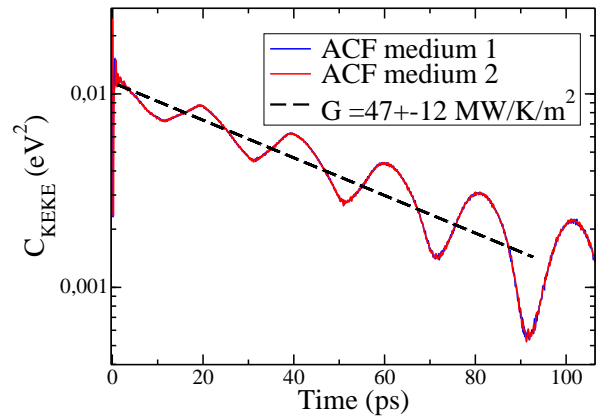}
\caption{(Color online) Kinetic energy autocorrelation function for the same system as fig. \ref{EACF}.}
\label{fig5}
\end{figure}

As we have discussed in the previous section, there are several formulae to compute the interfacial conductance from
EMD simulations probing the energy flux between the two solids. First, we will consider the energy auto-correlation formula in eq.~\ref{energy_Landauer}.  
To obtain the value of the Landauer conductance, one needs to compute the time derivative of the corresponding energy autocorrelation function. To this end, we have recorded the instantaneous value of the mechanical energy $E_i^m$ of each solid: 
\begin{equation}
\label{mechanical_energy}
E_i^m= \sum_{j \in i} \frac{1}{2}m \vec v_j^2 + \sum_{j,k \in i} V(\vec r_j- \vec r_k)
\end{equation}
where the first term represents the total kinetic energy of the solid $i$ and the second is the potential energy
between atoms belonging to the solid $i$. Note that this definition is somewhat a little bit arbitrary and we could have chosen 
to include in the mechanical energy the cross interaction term $\sum_{j \in i; k \in j} V(\vec r_j - \vec r_k)$. We have not observed significant differences in the value of the Landauer conductance as compared to the first definition. 
To determine the value of $G_{eq}$, we have computed the energy autocorrelation function (EACF) in each medium. 
The instantaneous value of the mechanical energy of each solid has been recorded every two time steps in the course of 
long NVE simulations corresponding to a total of $1$ million time steps.
The EACFs have been obtained by averaging over $10$ initial independent configurations.
An example of the averaged EACFs is shown in fig.~\ref{EACF}. The EACFs relative to the two solids are practically indistinguishable. 
Note the existence of very small oscillations of the EACFs at long correlation times. Given the value of the mean sound velocity in the system $c \simeq 1.2$ nm.ps$^{-1}$, these oscillations should probably correspond to long wavelength phonons which have travelled ballistically across the system several times, thus creating "echoes" in the correlation functions.
To estimate the value of the time derivative at time $t=0^{+}$ appearing in eq.~\ref{energy_Landauer}, we have fit the EACFs with a single exponential function between a time $t \sim 5$ ps and up to a time where the EACF has decreased by a factor $10$ as compared to the initial value. Using the fit $C_{EE}(t) = C_{EE}(0) \exp(-t/\tau)$, the conductance is 
$G = \frac{C_{EE}(0)}{2 \tau \mathcal A k_B T^2}$ where $C_{EE}(0)$ is found to be $k_B T^2 V \rho \bar c_v/4$ to a good approximation and the factor $2$ in $G$ comes from the fact that there are two interfaces due to the periodic boundary conditions.
The uncertainty in the determination of the value of $G$ is found to be typically $20$ percent for $10$ independent configurations 
and of course it decreases with the number of realizations of the system. 
Finally, we want to emphasize that we have observed that for large systems, the EACF decreases very slowly in good agreement 
with the previous mode analysis eq.~(\ref{relaxation_time}) which predicts that the mode relaxation times scale as the system length. \newline 
Alternatively, we have also analyzed the conductance using the Puech formula eq.~(\ref{Puech_formula})
where the instantaneous value of the flux may be estimated in the course of a MD simulation using the power of the interfacial forces~\cite{barrat2003}:
\begin{equation}
\label{flux_simulation}
q = \sum_{i \in 1; j \in 2} \vec v_i \cdot \vec F_{ij} 
\end{equation}
Note that this expression of the flux $q$ involves only atoms in the vicinity of the interface, while all the atoms of the system contribute to the expression based on eqs.~(\ref{energy_Landauer}) and (\ref{mechanical_energy}).
Figure~\ref{fig4} displays the running integral in the Puech formula eq.~(\ref{Puech_formula}) calculated using simulations 
for the same system considered in figure \ref{EACF}. The two curves correspond to the two interfaces of the system (remember the periodic boundary conditions). The running integrals display first a peak and then slowly decrease. Note the echoes in the upper curve.
We have found that it was difficult to define unambiguously a plateau eventhough we have considered here an average over $30$ independant configurations. The slow decrease has been also observed in the determination of liquid/solid conductance~\cite{barrat2003} and is predicted in eq.~(\ref{Green_Kubo_finite_system}). Indeed it is a common problem for a finite ergodic system that the Green Kubo formula predicts a vanishing transport coefficient~\cite{espanol1993} and in practice the running integral should be estimated at an intermediate time $\tau_0$ where the integral has not yet significantly decreased. The problem in heat transfer simulations of solids is that the spectrum of relaxation times $\tau_{\vec k}$ spans several decades and defining an intermediate time $\tau_0$ is not obvious in this situation. This difficulty is somewhat circumvent in the formula eq.~(\ref{energy_Landauer}) as it does not require to estimate a plateau.
\newline
Finally, we compare the value of the interfacial conductance to the expression proposed by Rajabpour and Volz~\cite{rajabpour2010}
for a classical system~:
\begin{equation}
\frac{1}{G} = \frac{1}{\mathcal{A} k_B} \int_{0}^{+\infty} \frac{\langle \delta T(t) \delta T(0) \rangle}{\langle \delta T(0)^2 \rangle} dt 
\left( \frac{1}{N_1} + \frac{1}{N_2} \right) 
\end{equation}
where $N_1$ and $N_2$ are the number of degrees of freedom characterizing each medium. Practically, the interfacial resistance 
is obtained by fitting the kinetic energy autocorrelation function with a single exponential having a decay time $\tau$.
The conductance is then given by $G=\mathcal A k_B \rho V /(2\tau)$.
Figure~\ref{fig5} displays the kinetic energy autocorrelation function obtained by averaging over $10$ independent simulations for the same system as considered before. As noted before \cite{rajabpour2010}, the kinetic energy 
displays a first fast decrease followed by a longer decrease, which is fitted with a single exponential with a relaxation time $\tau$. This latter time is used to obtain the interfacial conductance $G=\mathcal A k_B \rho V /(2\tau)$.
Note the oscillations in figure~\ref{fig5} due to the conversion between kinetic and potential energy. These oscillations occur with the same period than the period of echoes observed in the energy correlation function fig~\ref{EACF}.
The value of the interfacial conductance obtained $G=47 \pm 12$ MW/K/m$^2$ is smaller than the value obtained with the energy correlation function $G=60 \pm 12$ MW/K/m$^2$ but within the error bars. Hence, the two methods give consistent results and comparable error bars. \newline
We conclude by saying that compared to the Puech formula eq.~(\ref{Puech_formula}), the new Green-Kubo formula eq.~\ref{energy_Landauer} is easier to evaluate in a MD simulation because:
1. we do not need to estimate a plateau in a running integral; 2. the new formula involves all the atoms of the system while the Puech formula involves only atoms in the vicinity of the interface. As a result, the statistics is improved.

%
%
%
%

\section{Finite size effects \label{finite_size}}
\begin{figure}
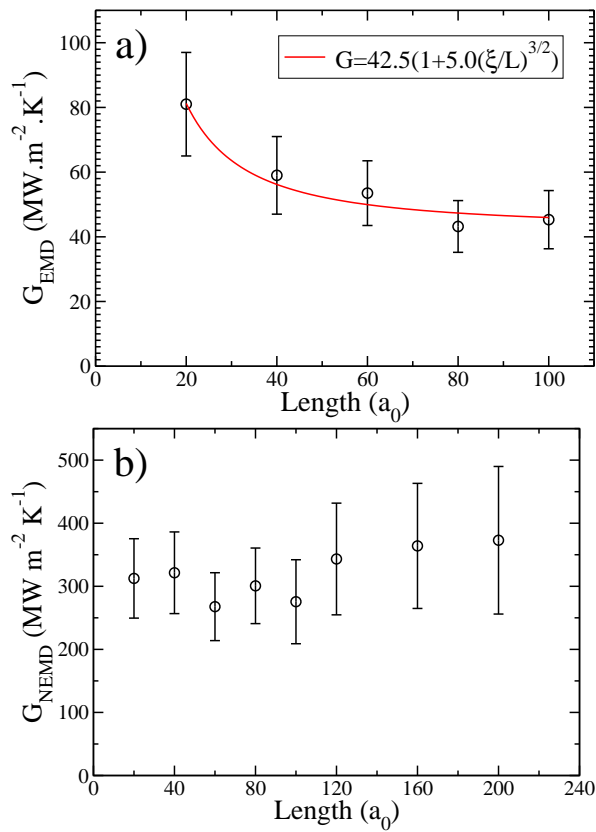

\includegraphics[width=0.9\linewidth]{Conductance_EMD_Length_T0.33.eps}
\includegraphics[width=0.9\linewidth]{Conductance_NEMD_Period.eps}
\caption{(Color online) 
 Interfacial conductance obtained by EMD (top figure) and NEMD (bottom figure) 
 as a function of the length of the system in units of the crystal monolayer $a_0$. The solid red line on the top figure displays the theoretical formula eq.~(\ref{conductance_length_theo}).
Same other parameters as fig. \ref{EACF}.}
\label{conductance_length}
\end{figure}

In this section, we compare the finite size effects in the determination of the conductance using both EMD and NEMD. Figure~\ref{conductance_length} a. displays the length dependance 
of the Kapitza conductance obtained with equilibrium simulations $G_{\rm EMD}$. It is found that $G_{\rm EMD}$ decreases with the system length. We have not studied the conductance of systems longer than $100$ $a_0$ because as explained above it leads to very long relaxation times $\tau_{\vec k}$ (see eq.\ref{relaxation_time}) and the determination of the equilibrium conductance becomes costly. In figure~\ref{conductance_length} b., we quantify the finite size effects on the conductance obtained with NEMD. The values obtained are consistent with the results of Stevens et al.\cite{stevens2007}. Note the values of the NEMD conductances which are larger than the EMD conductance by a factor $5$! This discrepancy will be analyzed in detail in the next section. We focus our attention here on the less severe size effects displayed by the NEMD conductance as compared with the EMD.
The finite size effects in the EMD method are quantitatively analyzed in the appendix~\ref{appx_finite_size}. 

\begin{figure}
\includegraphics[width=0.9\linewidth]{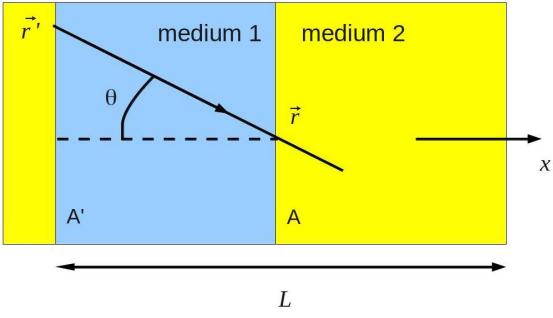}
\caption{(Color online)\label{illustration_cross_correlations} Postulated origin of the length dependance of the conductance $G_{\rm EMD}$ measured in EMD: long wavelength phonons may travel ballistically between the two interfaces $A$ and $A'$ thus creating thermal cross-correlations as measured by $\langle q_{A}(t)q_{A'}(0) \rangle$. 
This figure displays also the notations used in the appendix \ref{appx_finite_size} to quantify this effect.}
\end{figure}

The general idea is the following: in the EMD simulations, there are two interfaces between the two media to be considered because of the periodic boundary conditions as sketched in fig.~\ref{illustration_cross_correlations}.
These two interfaces are not necessarily independent from a thermal point of view: longwavelength phonons having a long mean free paths can create correlations between the instananeous value of the thermally induced flux at two adjacent interfaces. More precisely, if we denote by $A$ and $A'$ the two interfaces, the energy conservation writes:
$\frac{dE_1}{dt}=q_A+q_{A'}$ and the calculation of the equilibrium conductance involves cross terms of the form $\langle q_A(t)q_{A'}(0)\rangle$ and $\langle q_{A'}(t)q_A(0)\rangle$, while the "intrinsic" interfacial conductance is given by the term: $\langle q_A(t)q_A(0) \rangle =\langle q_{A'}(t)q_{A'}(0) \rangle$. Clearly, the cross terms will be relatively important at small interfacial separation $L/2$ because a majority of phonons modes will have a mean free path larger than $L$, while they should vanish in the limit $L \rightarrow \infty$. These cross terms are quantified in the appendix, under the assumptions of the interface between Debye solids with a constant transmission coefficient $t_{12}$, an assumption assessed {\it a posteriori} as shown in the next section \ref{mass_mismatch} where we will show that the EMD results are well described by the DMM model. We have also assumed that the phonon relaxation times are described by the Callaway model that we will discuss in the next section (cf eq.~\ref{callaway}). 
The prediction derived in the appendix~\ref{appx_finite_size} may be written:
\begin{equation} 
\label{conductance_length_theo}
G(L)=G_{\infty} \left(1 + c \left(\frac{\xi}{L}\right)^{3/2} \right) 
\end{equation}
where $G_{\infty}$ is the conductance caracterizing the interface between semi-infinite media, $c$ is a numerical constant and 
\begin{equation}
\xi = \frac{\lambda_i}{G_{ii}}
\end{equation}
is the phonon correlation length in medium i where $G_{ii}=\frac{3}{8} n_i k_B c_i$ is the EMD conductance between two identical media having the properties of medium $i$ (see also the next section). In principle, one should define two phonon correlation lengths characterizing the two media, but in the case of mass-mismatch Lennard-Jones solids, the correlation length is the same for the two media and is $\xi \simeq 6.3 a_0$ at $T=40$ K where we have used the value of the thermal conductivity obtained by Green-Kubo simulations by McGaughey and Kaviany~\cite{mcgaughey2004}.
Figure \ref{conductance_length} a. compares the EMD data to the theoretical expression eq.\ref{conductance_length_theo}
where we have obtained that the constant $c \simeq 5$ above the predicted value $c=2.75$ (see the appendix~\ref{appx_finite_size}).
The disagrement may be due to the use of the DMM model which as we will see in the next section tends to overestimate the
conductance obtained in EMD thus underestimating the constant $c$.
Note however that we could have fit with the same accuracy the EMD data using a functional form $G(L)=G_{\infty}(1+c(\xi/L))$ and that the $G(L) \sim L^{-3/2}$ scaling comes in our analysis from the Callaway assumption. What is important to remember is that the EMD conductance decays algebraically with the system length with a characteristic length proportional to the phonon correlation length $\xi$. On the other hand in NEMD the length dependence is smaller because the distribution of phonon mean free paths is cut due to the presence of heat resevoirs.

\section{Comparison between the EMD and NEMD conductances and theoretical models \label{conductance_series_massmismatch}}
In this section, we will compare the values of the interfacial conductance obtained either by EMD and NEMD (and extrapolated to infinite system lengths) to the expression of the conductance eqs.~(\ref{Landauer}) and (\ref{neq_conductance}) where we should specify the value of the phonon transmission coefficient $t_{12}$. To this end, we will consider two classical models for interfacial phonon scattering: the AMM and the DMM. We will generalize these two models to describe the non-equilibrium conductance~eq.(\ref{neq_conductance}). We will present in passing usefull approximate analytical expression 
to estimate both the Landauer conductance and the general conductance combined with the AMM model.   
\subsection{Debye approximation}
All along this section,we will make the assumption of Debye solids.
In the Debye approximation, the solids are assumed to have a constant group velocity which depends on the polarization mode
~\cite{ashcroft2002}. Most often an additional assumption is made consisting in assuming the same acoustic velocity 
for each polarization~\cite{dove1993}. For a three-dimensional crystal, this latter is defined as~:
\begin{equation}
\label{c_eff}
c_{\rm eff}=\frac{2c_T + c_L}{3}
\end{equation}
where the indexes $T$ and $L$ refer to the transverse and longitudinal polarizations respectively.
Under this assumption, the vDOS is~:
\begin{equation}
\label{Debye_vDOS}
g_{p}(\omega)=g(\omega)=\frac{\omega^2}{2\pi^2 c_{\rm eff}^3}
\end{equation} 
In the following we will drop the subscript "eff" and characterize the averaged sound velocity in medium $i$ by $c_i$.

\subsection{Acoustic Mismatch Model}
In the AMM, the phonons traveling towards the interface are assumed to see the interface as a sharp 
discontinuity of acoustic impedance $Z_i$ where 
\begin{equation}
\label{impedance}
Z_i = \rho_i c_i
\end{equation}
is given by the product of the mass density $\rho_i$ by the acoustic velocity in the medium i. As a result, phonons may be reflected by the interface or refracted on the other side of the interface following the equivalent of Snell laws~:
\begin{equation}
\label{Snell_law}
\frac{\sin \theta_1}{c_1} = \frac{\sin \theta_2}{c_2}
\end{equation} 
which strictly speaking holds as long as the incident angle is smaller than the critical angle  $\theta_c=\arcsin(c_2/c_1)$. Here $\theta_1$ and $\theta_2$ are the incident and refraction angles respectively. Above the critical angle, as for the electromagnetic waves, internal reflection occurs and the incident phonons are totally reflected. For the Si/Ge interface for which the ratio of the acoustic velocities is approximately $1.5$, the critical angle is $\theta_c \simeq 40 \deg$ and a significant fraction of phonons are totally reflected by the interface.
We have also assumed that a phonon conserves its polarization, i.e. there is no mode conversion and $c_1$ and $c_2$ denote
the acoustic velocities in media $1$ and $2$ corresponding to the same polarization state.
Another assumption behind eq.~\ref{Snell_law} is that the scattering is elastic, i.e. refracted and reflected phonons conserve their frequency. As a consequence, phonons having frequency 
above the Debye frequency of the softer solid are confined in the hard solid and not transmitted by the interface, i.e. 
the transmission coefficient is supposed to vanish. For phonons having frequencies smaller than the debye frequency of the softer solid, the transmission coefficient is derived from the Snell law~\cite{little1959} eq.\ref{Snell_law}: 
\begin{eqnarray}
\label{transmission_AMM}
t_{12}(\omega,\mu_1)& = & \frac{4Z_1Z_2\mu_1\mu_2}{(Z_1\mu_1+Z_2\mu_2)^2} \quad \omega < \min ({\omega_{D1},\omega_{D2}}) \nonumber \\
t_{12}(\omega,\mu_1)&= & 0 \; \mathrm{otherwise};
\end{eqnarray}
where we have introduced $\mu_i=\cos \theta_i$, and again it is implied that the incident angle is smaller than the critical angle.
At high temperatures, the regime relevant to classical molecular simulations where the equilibrium Bose-Einstein distribution $f_{\rm eq}(\omega) \rightarrow k_BT/\hbar \omega$, the AMM conductance which is calculated using the Landauer expression eq.~\ref{Landauer} may be written~:
\begin{equation}
\label{Landauer_AMM}
G_{\rm eq}^{\rm AMM} = \frac{3}{2} n_1 k_B c_1 \left(\frac{c_2}{c_1}\right)^{3} \int_{0}^{1} t_{12}(\mu_1) \mu_1 d \mu_1
\end{equation}
where $n_1$ denotes the number density of medium $1$. We have supposed without loss of generality that the medium denoted $2$
has the lowest Debye frequency. The factor $(\frac{c_2}{c_1})^3$ comes from the phonon confinement 
of high frequency phonons in medium $1$. The AMM conductance eq.~(\ref{Landauer_AMM}) should be evaluated numerically. Alternatively, one can obtain tractable analytical expressions for the AMM conductance, if we assume 
that when the acoustic contrast between the two solids is large, the transmission coefficient $t_{12}$ is dominated by phonons propagating with a small refraction angle, i.e. $\mu_2 \simeq 1$. Under this approximation, the AMM conductance is given by the approximate form:
\begin{equation}
\label{Landauer_AMM_appx}
G_{\rm eq}^{\rm AMM} \simeq G_{\rm eq}^{\rm AMM,appx} = \frac{3}{2} n_1 k_B c_1 \left(\frac{c_2}{c_1}\right)^{3} I_1^{\rm appx}
\end{equation}
where $I_1^{\rm appx}$ depends on the acoustic ratio $\beta=Z_2/Z_1$:
\small 
\begin{equation}
\label{I1_appx}
I_1^{\rm appx}=4\beta \left(1-\frac{\beta^2}{1+\beta}+\beta-2\beta\log \left(\frac{1+\beta}{\beta} \right)\right)
\end{equation}
\normalsize
As shown in the Appendix~\ref{appx_AMM}, the approximation  eq~(\ref{Landauer_AMM_appx}) gives a very good description of the AMM conductance over a wide range of acoustic contrast. 

\subsection{Diffuse Mismatch Model}
The previously described AMM model is supposed to predict the transmission of phonons of large wavelengths which behave as 
plane waves experiencing specular reflection or refraction at the interface. This model is commonly thought to apply at low temperatures where only long wavelength phonons are populated. At higher temperatures, interfacial scattering is thought to be diffuse like essentially because a majority of phonons have wavelengths comparable or even smaller than the interfacial roughness. This idea motivated the development of the DMM introduced by Swartz and Pohl~\cite{swartz1987,swartz1989} which 
 assumes that the phonons experiencing scattering at the interface loose totally the information about the medium where they come from. 
As a result, the probability that a phonon experiences a reflection in medium $2$ is equal to the probability that a phonon is transmitted from medium $1$ towards $2$~:
\begin{equation}
\label{amnesia}
t_{12}=1-t_{21}
\end{equation}
for the particular mode considered. Writing the total flux in medium $2$ 
together with the previous amnesia condition yields the transmission coefficient~:
\begin{equation}
\label{transmission_DMM}
t_{12}(\omega)=\frac{c_2 g_2(\omega)}{c_1 g_1(\omega) + c_2 g_2(\omega)}   
\end{equation}   
and as for the AMM model, it is implicitely assumed that high frequency phonons are confined in the harder material:
\begin{equation}
\label{confinement}
t_{12}(\omega)=0 \quad {\rm if} \quad \omega >\min(\omega_{D,1},\omega_{D,2})
\end{equation}
Since the transmission coefficient doesnot depend on the incident angle, the DMM conductance has a simple expression:
\begin{equation}
\label{Landauer_DMM}
G_{\rm eq}^{\rm DMM} = \frac{3}{4} n_1 k_B \frac{c_2^3}{c_1^2 + c_2^2} 
\end{equation}

\subsection{Generalized conductances}
To obtain tractable expressions for the non-equilibrium conductance eq.~(\ref{neq_conductance}) which depends on the fractions $\beta_{ij}$ eq.~(\ref{beta}), we need to do an hypothesis regarding the frequency-dependence of the phonon lifetime $\tau_i(\omega)$. The simplest is to assume that the phonon lifetime $\tau_i$ is controlled by Umklapp processes obeying Callaway model~\cite{callaway1959}:
\begin{equation}
\label{callaway}
\tau_{i}(\omega)=A_i \omega^{-2}
\end{equation}
where $A_i$ is a material parameter which depends on the temperature. Under this assumption and if interfacial scattering is supposed to be specular, the non-equilibrium conductance takes the form~: 
\small
\begin{equation}
\label{neq_conductance_AMM}
G_{\rm neq}^{\rm AMM} = \frac{G_{\rm eq}^{\rm AMM}}{1-\frac{3}{2}(\frac{c_2}{c_1})\left( \int_{0}^{1} \mu_1^2 t_{12}(\mu_1) d\mu_1 + \frac{c_2}{c_1} 
\int_{0}^{1} \mu_1 \mu_2 t_{12}(\mu_1) d \mu_1 \right)}
\end{equation}
\normalsize
where $\mu_2$ denotes the cosine of the refracted angle~\cite{note}:$\mu_2=\sqrt{1-(c_2/c_1)^2(1-\mu_1^2)}$. Again, the conductance eq.~(\ref{neq_conductance_AMM}) can be approximated:
\begin{equation}
\label{neq_conductance_AMM_appx}
G_{\rm neq}^{\rm AMM} \simeq \frac{G_{\rm eq}^{\rm AMM,appx}}{1-\frac{3}{2}(\frac{c_2}{c_1})\left( I_2^{\rm appx} + \frac{c_2}{c_1} 
I_3^{\rm appx} \right)}
\end{equation}
where $I_3^{\rm appx}=I_1^{\rm appx}$ is defined in eq.(\ref{I1_appx}) and~:
\small 
\begin{equation}
\label{I2_appx}
I_2^{\rm appx}=4\beta \left(\frac{1}{2}-2\beta-\beta^2+\frac{\beta^3}{(1+\beta)}+3\beta^2\log \left(\frac{1+\beta}{\beta}\right) \right)
\end{equation}
\normalsize
The accuracy of the approximation eq~(\ref{neq_conductance_AMM_appx}) and a finer approximation are presented in the Appendix~\ref{appx_AMM}.
The conductance obtained using the DMM transmission coefficient is~:
\begin{equation}
\label{neq_conductance_DMM}
G_{\rm neq}^{\rm DMM} = \frac{3}{4} n_1 k_B \frac{c_2^3}{c_1^2 + (c_1-c_2)^2} 
\end{equation}
Again we note that when the two media are similar, $c_1=c_2$ and the previous equation for the conductance predicts a finite conductance $G(c_1=c_2)=\frac{3}{4} n_1 k_B c_1$. This new paradox can be traced back to the use of the DMM transmission coefficient eq.~(\ref{transmission_DMM}) which tends towards $1/2$ when $c_1 \rightarrow c_1$. This problem disappears using the AMM transmission coefficient 
because the denominator of eq.~(\ref{neq_conductance_AMM}) tends towards $0$ when the two media are identical.

\subsection{Interfacial conductance of a series of mass-mismatched Lennard-Jones solids \label{mass_mismatch}}
\begin{figure}
\includegraphics[width=1.1\linewidth]{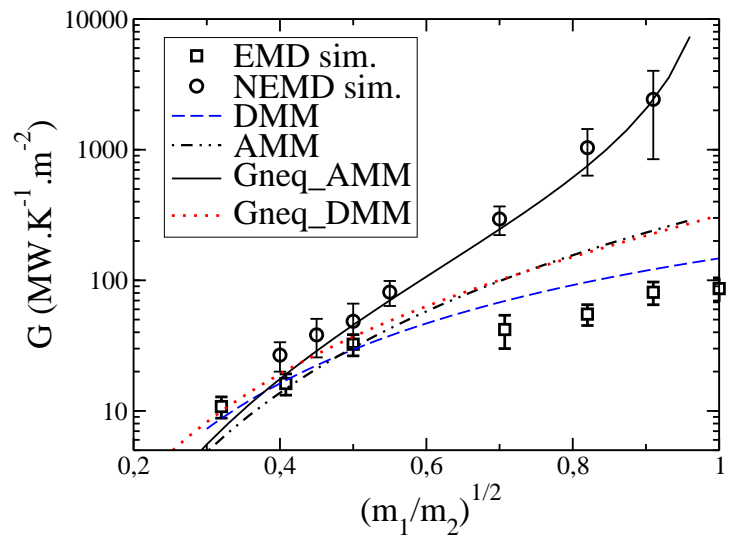}
\caption{(Color online) Interfacial conductance determined by EMD and NEMD as a function of the mass mismatch between the two solids. The simulations results are compared with the different theoretical expression AMM (eq~\ref{Landauer_AMM}), DMM (eq~\ref{Landauer_DMM}) and the generalizations eqs.~(\ref{neq_conductance_AMM}) and (\ref{neq_conductance_DMM}). The temperature is $T=40$ K.}
\label{conductance_mass}
\end{figure}

In this subsection, we compare the conductances obtained using both EMD and NEMD simulations to the previous equations for the interfacial conductance, respectively given by the AMM model eqs.~(\ref{Landauer_AMM}), the DMM model eq.~(\ref{Landauer_DMM}) and the generalizations eqs.~(\ref{neq_conductance_AMM}), and eq.~(\ref{neq_conductance_DMM}). In evaluating these different expressions 
for the case of the interface between Lennard-Jones solids, we have used the values of Argon: $c_1=1250$ m.s$^{-1}$ for the average sound velocity of the harder medium and a number density $n=2.57$ $10^{28}$ m$^{-1}$.
\newline
In figure~\ref{conductance_mass}, we have reported the values obtained using EMD and NEMD simulations for the interfacial conductance characterizing the interface between LJ solids having a variable mass ratio. This ratio has been varied between $1$ and $10$ so as to change the acoustic impedance ratio $Z_1/Z_2=\sqrt{m_1/m_2}$ between the two media 
between $1$ and $0.3$. The EMD values have been obtained using the finite size scaling analysis described before and the extrapolation to infinite system length as described in the previous section~\ref{finite_size}. On the other hand the NEMD values have been obtained using a total system length of $200$ $a_0$.
The trend displayed by the NEMD data is very similar to the NEMD simulation results of Landry and McGaughey for the Si/heavy Si interface~\cite{landry2009}.
Strikingly and as already in the previous section, the EMD and NEMD values may differ significantly depending on the acoustic contrast between the two solids. In particular, when the dissimilarity between the two solids is small, the NEMD conductance is larger than the EMD value by more than one order of magnitude ! Note that the corresponding impedance ratio $\simeq 1.5$ are typical of AlAs/GaAs interface~\cite{termentzidis2010}. Even for dissimilar solids like Si/Ge for which the impedance ratio is $\simeq 1.7$, the difference may reach a factor $3$! This discrepancy may be simply explained: as we showed, the EMD conductance yields the Landauer expression of the conductance eq.~(\ref{Landauer}) while the NEMD value should be akin to the Simon conductance eq.~(\ref{neq_conductance}). The difference between the two values of the conductance is quantified by the fractions of "out-of-equilibrium" phonons $\beta_{12}$ and $\beta_{21}$ (eqs.~\ref{beta}) which tend to make the denominator of eq.~(\ref{neq_conductance}) vanishing when the acoustic properties of the two solids become comparable. In this limit, the difference between the general expression eq.~(\ref{neq_conductance}) and the Landauer conductance may be very large, yielding the divergence of the NEMD conductance when the two solids are similar. In figure \ref{conductance_mass}, we have also compared the EMD values to the AMM and DMM models which are consistent with Landauer formalism. 
Based on the analysis of the conductance at the interface between similar solids,
we conclude that the DMM model gives a relatively good description of the EMD conductance, while the AMM model overpredicts the EMD values by a factor $2$. Note however that the difference between the AMM and DMM models is not that large for dissimilar materials. The small discrepancy between the simulation values and the DMM model may come from our assumption of Debye solids in a situation where a fine description of high frequency modes is required, as the DOS of the two solids strongly overlap and the maximal frequency transmitted by the interface $\omega_{\rm max}$ tends towards the Debye frequency of the harder solid. Regarding the NEMD values, it is clear that the generalization eq.~(\ref{neq_conductance_AMM}) based on the acoustic transmission coefficient describes quite satisfactorily the divergence of the NEMD conductance. Equation~(\ref{neq_conductance_DMM}) which relies on a diffusive transmission coefficient underpredicts the NEMD conductance by a factor larger than $5$ for typical values of the acoustic impedance ratio. This is not completely surprising since as we discussed before, if interfacial scattering is diffuse, the interfacial conductance does not diverge when the two solids become similar. Also importantly, we have seen that interfacial phonon transmission in EMD simulations is controlled by diffuse events, while it becomes determined by the acoustic properties of the two solids when a thermal flux is imposed. Hence, we conclude that the energy transmission coefficient is not an intrinsic property of an interface, and it may depend on the nature of the source of thermal flux (i.e. external heat reservoirs vs. internal fluctuations). Given the results of the simulations, we are tempted to conclude that in equilibrium simulations, thermal fluctuations destroy the correlations between incident and transmitted phonons so that the amnesia condition eq.~(\ref{amnesia}) is verified and the conductance is well predicted by the DMM model. In particular when the two media are similar, one recovers the fact that a phonon in excess will have a probability $1/2$ to be transmitted and $1/2$ to be reflected, which is consistent with the EMD values obtained in this limit. 
On the other hand, in a NEMD simulation the situation is quite different: indeed phonons travelling across the interface see the interface as a sharp discontinuity which creates strong correlations between incident and transmitted phonons. Because the thickness of the interface is smaller than the phonon wavelengths, the transmission and reflection coefficients will be in these conditions controlled by the acoustic impedances of the two media, and in the limit of similar solids the transmission coefficient should approach $1$. This may explain the difference in transmission coefficients between EMD and NEMD simulations.

\section{Conclusion}
In conclusion, we have analyzed two methods to measure the thermal Kapitza conductance between dielectrics using molecular dynamics. We have proposed a new Green-Kubo formula (eq.~\ref{energy_Landauer}) to measure the interfacial conductance using equilibrium EMD simulations. This formula is easier to evaluate in a molecular dynamics simulations 
as compared to the classical formula eq.~(\ref{Puech_formula}) because it avoids to estimate a plateau in the running integral of a correlation function. Also the statistics is improved because the new formula involves all the atoms of the system when 
the Puech formula considers only the atoms in the vicinity of the interface.
We have also analyzed finite size effects in EMD and showed that their origin is the correlation between 
the interfaces created by long wavelength phonons which travel ballistically across the periodic simulation cell. 
On the other hand, in NEMD the distribution of phonon mean free paths is cut due to the presence of the heat reservoirs. 
This effect explains why finite size effects are less severe in NEMD than in EMD. We have also shown that the interfacial conductance measured in an EMD simulation whether using the Puech formula or the energy correlation function identifies with the Landauer conductance which assumes phonons on both sides of the interface to be at equilibrium. This explains why in EMD a finite conductance is measured when the two solids are similar. On the other hand, we have explained that in NEMD simulations, we measure a conductance given by the general expression eq.~(\ref{neq_conductance}) inspired by Simons, and which accounts for the out-of-equilibrium distribution of phonons consistent with the imposed heat flux. Hence, we conclude that the two methods give {\em intrinsically} different values of the interfacial conductance. For impedance ratios typical of real interfaces, the difference in conductances is large corresponding typically to a factor between $4$ and $10$. On the other hand, when the impedance ratio is large the difference in conductances is small. This explains why Barrat found good agreement between EMD and NEMD in the case of solid/liquid interfaces.
Also we have shown that the two methods probe different energy transmission coefficients: EMD conductance are consistent with transmission describing diffuse events whose rates are governed primarily by the density of states mismatch between the two solids. On the other hand, in NEMD the transmission of phonons probed is specular in nature at least in the case analyzed here of atomically perfect interfaces. This difference stems from the different origin of the flux instantaneously flowing across the interface. \newline 
An important question that we have to answer is which technique should be used-NEMD or EMD-to access a conductance measured experimentally. Intuitively, NEMD should be used to compute the value of the conductance measured experimentally using steady state technique such as the $3$ omega method. On the other hand, EMD should be more akin to Laser pump-probe experiments where the transient response to an initial heating is recorded~\cite{juve2009}. This needs further theoretical analysis and will be the subject of future investigation. \newline    
Another interesting question deals with the role of ballistic phonons in the derivation of the non-equilibrium conductance eq.~(\ref{neq_conductance}). Indeed, Landry and McGaughey observed that the non-equilibrium conductance overestimates the conductance measured at the interface between Si and Ge. We think that this discrepancy stems from the large value of the dominant mean free path in Si which is comparable with the system size considered. \newline
All these results have been obtained for the case of the perfect interfaces. This may allow to disentangle effects related to the contrast between the vibrational properties of the {\em bulk} media from the effect arising from the interface. In particular, at the interface between real materials eventhough the interface may be treated to become atomically sharp, there is always a lattice mismatch which may enhance diffuse phonon scattering. 
The use of MD models allows then to measure each effect separately thus opening the way to a fundamental understanding of interfacial heat transfer between solids.

\section{Appendix: Finite size effects in the determination of the EMD conductance \label{appx_finite_size}}
In this appendix, we derive the length-dependent conductance eq.~\ref{conductance_length_theo} measured in the EMD simulation. Please refer to the figure \ref{illustration_cross_correlations} for the relevant notations to be used here.
As explained in the main body of the text, the length dependance of the conductance measured in EMD simulations is assumed to be caused by cross-correlations between the fluxes across the two interfaces of the system. We focus then on the cross correlation term $\langle q_{\rm A}(t) q_{\rm A'}(0) \rangle$. We write the interfacial fluxes in terms of the phonons distribution functions in medium $i$: $f_{i,\vec k}(\vec r,t)=f_{i,\vec k}^{\rm eq}(\vec r) + \delta f_{i,\vec k}(\vec r,t)$ where we have omitted the index designating the polarization to simplify the dicusssion. Hence, we have :
\small
\begin{eqnarray}
q_{\rm A'}(t=0)&=&\int_{\rm A'} \frac{1}{V} \sum_{v_{2x}>0} t_{21} \delta f_{2,\vec k}(0,\vec r^{'}_{//},t=0) \hbar \omega v_{2x} d {\vec r^{'}}_{//} \nonumber \\ 
+\int_{\rm A'} &&\frac{1}{V} \sum_{v_{1x}<0} t_{12} \delta f_{1,\vec k}(0,{\vec r^{'}}_{//},t=0) \hbar \omega v_{1x} d {\vec r^{'}}_{//}
\end{eqnarray}
\normalsize
A similar equation holds for $q_{\rm A(t)}$ but in the following, we will use the following equation which derives from the continuity of the interfacial flux:
\begin{eqnarray}
q_{\rm A}(t)&=&\int_{\rm A} \frac{1}{V} \sum_{\vec k} \delta f_{2,\vec k}(L/2,\vec r_{//},t) \hbar \omega v_{2x} d \vec r_{//} \nonumber \\ 
&=&\int_{\rm A} \frac{1}{V} \sum_{\vec k} \delta f_{1,\vec k}(L/2,\vec r_{//},t) \hbar \omega v_{1x} d \vec r_{//}
\end{eqnarray}
The cross correlation $\langle q_{\rm A}(t) q_{\rm A'}(0) \rangle$ will thus involve correlation of the phonon distribution function of the form $\langle \delta f_{i,\vec k}(0,\vec r,t=0) \delta f_{i,\vec k'}(L/2,\vec r^{'},t) \rangle$. We assume the thermally induced 
phonon modes propagating in the medium $i$ to be incoherent and characterized by a mean free path $\Lambda_{i,\vec k}$. Under these conditions, the phonon correlation writes:
\begin{eqnarray}
\langle \delta f_{i,\vec k}(\vec r^{'},t) \delta f_{i,\vec k'}(\vec r,0) \rangle = \nonumber \\ 
V\langle \delta f_{2,\vec k}^2 \rangle \delta (\vec r^{'}-\vec r-\vec v_{i} t) \exp \left(-\frac {\vert \vec v_{i} t \vert}{\Lambda_{i, \vec k}} \right) \delta_{\vec k,\vec k'}
\end{eqnarray}  
where 
\begin{equation}
\langle \delta f_{i,\vec k}^2 \rangle = \frac{{\bar c}_v k_BT^2}{2V(\hbar \omega)^2} 
\end{equation}
The cross correlation flux writes then: 
\small
\begin{equation}
\langle q_{\rm A'}(t) q_{\rm A}(0) \rangle = \sum_{v_{2x}>0} \frac{\mathcal {A} k_B T^2 {\bar c}_v}{2V} t_{21} v_{2x}^2 \delta \left(\frac{L}{2}-v_{2x}t \right) 
\exp \left(-\vert \vec v_2 \vert t/\Lambda_2 \right)
\end{equation}
\normalsize
and the contribution to the conductance is 
\small
\begin{equation}
\frac{1}{\mathcal {A} k_B T^2} \int_{0}^{+\infty} \langle q_{\rm A'}(t) q_{\rm A}(0) \rangle dt = \sum_{v_{2x}>0} \frac{{\bar c}_v t_{21} v_{2x}}{2V} \exp \left(-L/2 \cos \theta \Lambda_2 \right)
\end{equation}
\normalsize
To evaluate this latter conductance, we transform the discrete sum in a integral over the frequency:
\begin{eqnarray}
\label{conductance_I}
\frac{1}{\mathcal {A} k_B T^2} \int_{0}^{+\infty} \langle q_{\rm A'}(t) q_{\rm A}(0) \rangle dt \nonumber \\
= \frac {3}{4} \int_{0}^{\omega_{\rm max}}g_2(\omega) {\bar c}_v t_{21}(\omega) \vert v_2 \vert I_2 (L,\omega) d\omega 
\end{eqnarray}
where we have supposed that the transmission coefficient $t_{21}$ is independent on the incidence angle and we have introduced the 
integral:
\begin{equation}
I_2(L,\omega) = \int_{1}^{+\infty} u^{-3} \exp \left(-\frac{Lu}{2 \Lambda_2(\omega)}\right) du
\end{equation} 
For thick media, $L \gg \Lambda$ and one can approximate the integral $I(L,\omega) \sim \frac{2\Lambda}{L} \exp \left(-\frac{L}{2 \Lambda} \right)$~\cite{modest2003}. To evaluate the conductance eq.~(\ref{conductance_I}), we assume as stated in the main body of the text that the vDOS is described by the Debye model and the frequency dependence of the mean free path is given by Callaway law~\cite{callaway1959}:
\begin{equation}
\Lambda_2(\omega)=A_2 \vert v_2 \vert / \omega^2 
\end{equation}
where $v_2$ is assumed to be constant consistently with our hypothesis of Debye solid.
The constant $A_2$ is related to the thermal conductivity $\lambda_2$ through:
\begin{equation}
\lambda_2 = \frac{k_B A_2 \omega_{D,2}}{2 \pi^2 \vert v_2 \vert}
\end{equation}
If we suppose furthermore that the transmission coefficient $t_{12}$ is independent on the frequency $\omega$ as in the DMM model, it comes:
\small
\begin{equation}
\frac{1}{\mathcal {A} k_B T^2} \int_{0}^{+\infty} \langle q_{\rm A'}(t) q_{\rm A}(0) \rangle dt \nonumber \\
= \frac {3}{8} k_B n_2 t_{21} v_2 \left( \frac{2^{1/3} A_2}{n_2^{2/3} \pi v_2 L} \right)^{3/2} 
\end{equation}
\normalsize
where we have assumed that $\omega_{\rm max} \gg \sqrt{v_2 A_2/L}$, which physically means that the phonon mean free path of the mode with a frequency 
$\omega_{\rm max}$ is smaller than the system length.
This latter contribution may be rewritten:
\begin{equation}
\delta G(L) = \frac{3 n_2 k_B t_{21} v_2}{8} \sqrt{\frac{9 \pi}{8}} \left( \frac{\lambda_2}{G_{22} L} \right)^{3/2}
\end{equation}
where we have introduced the conductance $G_{22}=3n_2 k_B v_2/8$. Similar calculations allow to express the total contribution of the cross fluxes as:
\footnotesize 
\begin{equation}
\delta G(L) = \frac{3}{8}k_B \sqrt{\frac{9 \pi}{8}} \left(n_2 t_{21} v_2 \left(\frac{\lambda_2}{G_{22} L} \right)^{3/2} + n_1 t_{12} v_1 \left(\frac{\lambda_1}{G_{11} L} \right)^{3/2} \right)
\end{equation}
\normalsize
Note that for Lennard-Jones solids differing only by their mass, the length $\xi=\lambda_i/G_{ii}=A_i/(n_i^{2/3}v_i)$ is constant independent on the mass. Again anticipating the results of the section~\ref{mass_mismatch}, we can assume that the infinite length conductance $G_{12}$ is given by $G_{12}=\frac{3}{4}k_B n_2 v_2 t_{21}$ and the transmission coefficient obeys: 
$t_{12}=m_2 t_{21}/m_1$ where we have introduced the masses of the two solids. Hence for the interface considered in fig.~\ref{EACF}
for which $m_2/m_1=2$, the correction to the conductance writes:
\begin{equation}
\delta G(L)=c G_{12} \left( \frac{\xi}{L} \right)^{3/2}
\end{equation}  
where $c=\sqrt{\frac{\pi}{6}}(1+2\sqrt{2}) \simeq 2.75$.

\begin{figure}
\includegraphics[width=0.9\linewidth]{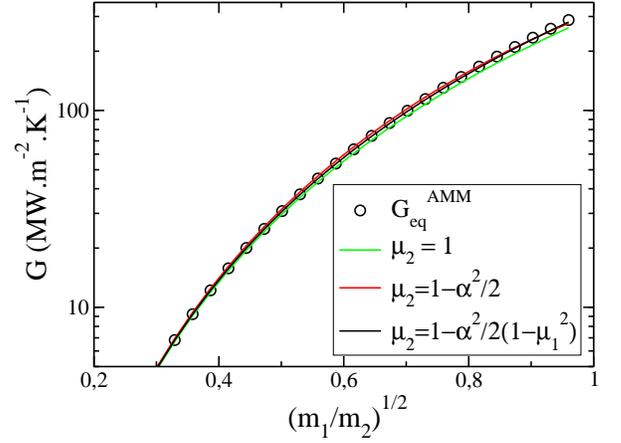}
\caption{(Color online) Comparison between the exact Landauer AMM conductance eq.~\ref{Landauer_AMM} with the approximate solution
eq.~(\ref{Landauer_AMM_appx}) for the series of mass-mismatched Lennard-Jones crystals considered in the simulations. The mean acoustic velocity is taken to be that of the Lennard-Jones Argon.}
\label{fig10}
\end{figure}

\begin{figure}
\includegraphics[width=0.9\linewidth]{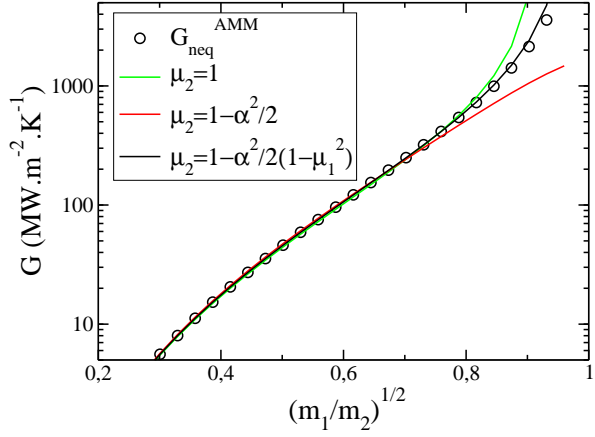}
\caption{(Color online) Comparison between the exact non-equilibrium conductance eq.~(\ref{neq_conductance_AMM}) with the approximate solution eq.~(\ref{neq_conductance_AMM_appx}) for the series of mass-mismatched Lennard-Jones crystals considered in the simulations. The mean acoustic velocity is taken to be that of the Lennard-Jones Argon.}
\label{fig11}
\end{figure}

\section{Appendix: approximations of the AMM conductances \label{appx_AMM}}
In this appendix, we assess the accuracy of different approximations used to estimate the conductance appearing in eqs.~(\ref{Landauer_AMM}) and~(\ref{neq_conductance_AMM}).
More specifically, one needs to approximate the three geometrical integrals which depend on the Rayleigh transmission coefficient eq~\ref{transmission_AMM}:
\begin{eqnarray}
\label{integrals}
I_1 & = &\int_{0}^{1} \mu_1 t_{12}(\mu_1) d\mu_1;  \\
I_2 & = &\int_{0}^{1} \mu_1^2 t_{12}(\mu_1) d\mu_1 \\
I_3 & = &\int_{0}^{1} \mu_1 \mu_2 t_{12}(\mu_1) d\mu_1 
\end{eqnarray}
The general idea is to assume that the geometric integrals are dominated by phonons propagating in the soft material with a small angle $\theta_2 \simeq 0$ when the acoustic mismatch betwen 
the two solids is large. This leads to use the following approximations:
\begin{eqnarray}
\mu_2 &\simeq& 1 \label{appx_1}\\
\mu_2 &\simeq& 1-\frac{\alpha^2}{2} \label{appx_2}\\
\mu_2 &\simeq& 1-\frac{\alpha^2}{2}(1-\mu_1^2) \label{appx_3} 
\end{eqnarray}
where we have denoted by $\alpha$ the ratio of the sound velocities~:~$\alpha=c_2/c_1<1$.
The first approximation eq.~(\ref{appx_1}) has been already discussed in the text and the corresponding approximate integrals 
are given in eqs.~(\ref{I1_appx}) and (\ref{I2_appx}). In the second approximation eq.~(\ref{appx_2}), the three approximated integrals 
depend on the parameter $\gamma=\beta(1-\frac{\alpha^2}{2})$:
\small
\begin{eqnarray}
\label{integrals_appx}
I_1^{\rm appx}&=&4\gamma \left(1-\frac{\gamma^2}{1+\gamma}+\gamma-2\gamma\log \left(\frac{1+\gamma}{\gamma} \right)\right) \nonumber \\
I_2^{\rm appx}&=&4\gamma \left(\frac{1}{2}-2\gamma-\gamma^2+\frac{\gamma^3}{(1+\gamma)}+3\gamma^2\log \left(\frac{1+\gamma}{\gamma}\right) \right) \\
I_3^{\rm appx}&=&\left(1-\frac{\alpha^2}{2}\right)I_1^{\rm appx}
\end{eqnarray}
\normalsize
The third approximation eq.~(\ref{appx_3}) yields calculations a little bit more involved.
Within this approximation, one obtains the following expressions for the three integrals:
\begin{eqnarray}
I_1^{\rm appx}=4\beta \int_{0}^{1} \frac{\mu^2 (1-\frac{\alpha^2}{2}(1-\mu^2))}{(\mu+\beta(1-\frac{\alpha^2}{2}(1-\mu^2)))^2} d\mu;  \\
I_2^{\rm appx}=4\beta \int_{0}^{1} \frac{\mu^3 (1-\frac{\alpha^2}{2}(1-\mu^2))}{(\mu+\beta(1-\frac{\alpha^2}{2}(1-\mu^2)))^2} d\mu;  \\
I_3^{\rm appx}=4\beta \int_{0}^{1} \frac{\mu^2 (1-\frac{\alpha^2}{2}(1-\mu^2))^2}{(\mu+\beta(1-\frac{\alpha^2}{2}(1-\mu^2)))^2} d\mu;  
\end{eqnarray}
The denominator appearing in the three integrals has two poles $r_{1/2}$ having multiplicity two and which  are given by~:
\begin{eqnarray}
r_1 = \frac{-1+\sqrt{1-2\alpha^2\beta²(1-\frac{\alpha^2}{2})}}{\alpha^2 \beta} \\
r_2 = \frac{-1-\sqrt{1-2\alpha^2\beta^2(1-\frac{\alpha^2}{2})}}{\alpha^2 \beta} 
\end{eqnarray}
and the approximated integral $I_j^{\rm appx}$ are given by~:
\begin{widetext}
\begin{equation}
\label{appx_integrals}
I_j^{\rm appx} = 4\beta \left( E_j + \frac{a_{1j}}{r_1(r_1-1)}
 + b_{1j} \log((1 - r_1)/(-r_1)) + \frac{a_{2j}}{r_2(r_2-1)}
 + b_{2j} \log((1 - r_2)/(-r_2)) \right) \quad j \in {1,2,3} 
\end{equation}
\end{widetext}
with~:
\begin{eqnarray}
E_1 & = &  \frac{2} {\beta^2 \alpha^2}  \\
E_2 & = &  \frac{1} {\beta^2 \alpha^2} - \frac{8}{\beta^3 \alpha^4}  \\
E_3 & = &  \frac{1} {3\beta^2}-\frac{2}{\beta^3 \alpha^2}+\frac{12}{\beta^5 \alpha^4} 
\end{eqnarray}
and~:
\begin{widetext}
\begin{eqnarray}
a_{i1} & = & -\frac{4{r_i}^3}{\beta^3 \alpha^4 (r_1-r_2)^2} \\
b_{i1} & = & \frac{r_i(1-r_i)}{r_i-r_{i+1}} \left( \frac{1-r_{i+1}}{(1+\beta)^2} -\frac{2}{\beta^2 \alpha^2} 
-a_{i1} \left(\frac{1-r_{i+1}}{(1-r_i)^2} + \frac{r_{i+1}}{{r_i}^2}\right) - \frac{a_{i+1,1}}{r_{i+1}(1-r_{i+1})}
 \right) \\
a_{i2} & = & -\frac{4{r_i}^4}{\beta^3 \alpha^4 (r_1-r_2)^2} \\
b_{i2} & = & \frac{r_i(1-r_i)}{r_i-r_{i+1}} \left(  \frac{1-r_{i+1}}{(1+\beta)^2} 
+\frac{8}{\beta^3\alpha^4} -\frac{2(1-r_{i+1})}{\beta^2 \alpha^2} 
-a_{i2} \left(\frac{1-r_{i+1}}{(1-r_i)^2} + \frac{r_{i+1}}{{r_i}^2}\right) -\frac{a_{i+1,2}}{r_{i+1}(1-r_{i+1})} \right) \\
a_{i3} & = & \frac{4{r_i}^4}{\beta^4 \alpha^4 (r_1-r_2)^2} \\
b_{i3} & = & \frac{r_i(1-r_i)}{r_i-r_{i+1}} \left( \frac{1-r_{i+1}}{(1+\beta)^2}
-\frac{(\beta\alpha^2-4)(1-r_{i+1})}{\beta^3 \alpha^2}-\frac{12}{\beta^4 \alpha^4} 
-a_{i3} \left(\frac{1-r_{i+1}}{(1-r_i)^2} + \frac{r_{i+1}}{{r_i}^2}\right) -\frac{a_{i+1,3}}{r_{i+1}(1-r_{i+1})} \right)
\end{eqnarray}
\end{widetext}
where $i\in {1,2}$ and we have noted $i+1 = 1 +(i \mod(2))$, i.e, $1+1 \equiv 2; 2+1 \equiv 1$.
We can now study the accuracy of the previous approximations by comparing the exact expressions for the conductance 
eqs.~(\ref{Landauer_AMM}) and (\ref{neq_conductance_AMM}) with the approximate equations involving the three approximations discussed above.
 The comparison is shown in the figures~(\ref{fig10}) and (\ref{fig11}) for the series of mass mismatched Lennard-Jones solids analyzed in the simulations. 
Strikingly the different approximations seem to work quite well over a broad range of acoustic impendance ratios. The first approximation eq~(\ref{appx_1}) slighlty underestimates the Landauer AMM conductance when the impedance ratio tends towards $1$, but the two other approximations describe accurately the Landauer conductance 
for the whole range of ratio. As for the non-equilibrium conductance, the three approximations work quite well when the impedance ratio is smaller than $0.8$.
Above $0.8$, the approximations eqs~(\ref{appx_1}) and (\ref{appx_2}) respectively overestimate and underestimate the conductance. In particular, eq.~(\ref{appx_1})
predicts the divergence of the conductance at a value of the impedance ratio $<1$.
 On the other hand the approximation eq~(\ref{appx_3}) predicts accurately the final divergence of the conductance up to ratios $\sim 0.9$.  None of the approximation presented predicts the divergence of the conductance when the impedance ratio tends towards $1$, but in practice it is not common to work with such large ratios.

\begin{acknowledgments}
Simulations have been run at the "Pole Scientifique de Mod\'elisation Num\'erique" de Lyon using the LAMMPS open source package~\cite{plimpton1995}. We acknowledge interesting discussions with P. Chantrenne, T. Albaret, J.-Y. Duquesne and S. Volz.
\end{acknowledgments}


\begin{thebibliography}{99}
\bibitem{note_infinite_conductance} A comment is of interest here. Indeed, the interfacial quantity between two media should be evaluated at the 
nominal position of the Gibbs dividing surface, typically at half distance between two atomic layers. Hence, the temperature profile should be extrapolated 
in each medium to the position of this surface. In this case, one measures a vanishing temperature drop and an infinite conductance at the interface between 
two similar media. On the other hand, if one estimates a temperature jump between two adjacent monolayers, one obtains an interfacial conductance on the order of 
$\lambda/a$ where $\lambda$ is the bulk conductivity of the material considered and $a$ is the distance between nearest neighbouring atomic layers. This value of the conductance is typically two orders of magnitude larger than the maximal conductance predicted by the Landauer equation in the classical limit $n k_B v$ where 
$n$ is the number density and $v$ is the solid acoustic velocity.
\bibitem{note2} Indeed this assumption is not necessary. Eq.~\ref{steady_state_distribution} can be shown to be valid when the characteristic length 
of the problem (the distance between the thermostats) is larger than the phonon mean free path.
\bibitem{note_derivative} The derivative of $f(t)=\exp(-|t|/\tau_{\vec k})$ may be calculated using 
$f(t)=H(t)\exp(-t/\tau_{\vec k})+H(-t)\exp(t/\tau_{\vec k})$
where $H(t)$ is the Heaviside function. Using $H'(t)=\delta(t)$ one arrives at $f'(t)=(\text{sgn}(t)/\tau_{\vec k})\exp(-|t|/\tau_{\vec k}) +2\delta(t)$. Note that indeed the derivative should be understand here in terms of generalized functions (distributions)
\bibitem{note_thermodynamic_limit} This scaling is obtained because there are $\propto V$ terms in the sum over the allowded modes and 
because $\tau_{\vec k} \propto V/\mathcal A$.
The thermodynamic limit in our case means $\mathcal A \rightarrow \infty$, $V \rightarrow \infty$ and 
$V/\mathcal A \rightarrow \infty$. As usual with Green-Kubo formulae, it is important to consider first the limit 
$V \rightarrow \infty$ before the limit $t \rightarrow \infty$.
\bibitem{note} To arrive at eq.~\ref{neq_conductance_AMM}, we have also used $\int_{0}^{\sqrt{1-(c_2/c_1)^2}} \mu_2^2 t_{21}(\mu_2) d \mu_2 = (\frac{c_2}{c_1})^2 \int_{0}^{1} \mu_1 \mu_2 t_{12}(\mu_1) d \mu_1$.
\end{thebibliography}
\end{document}